\documentclass[aps,prb,reprint,superscriptaddress]{revtex4-1}
\usepackage{textcomp}
\usepackage{graphicx} 
\usepackage{epsfig}
\usepackage{epstopdf}
\usepackage{longtable}
\usepackage{amsmath}

\begin{document}

\title{Coincident structural and magnetic order in BaFe$_2$(As$_{1-x}$P$_x$)$_2$ revealed by high-resolution neutron diffraction}

\author{J. M. Allred}
\affiliation{Materials Science Division, Argonne National Laboratory, Argonne, IL 60439-4845, USA}
\author{K.M. Taddei}
\affiliation{Materials Science Division, Argonne National Laboratory, Argonne, IL 60439-4845, USA}
\affiliation{Physics Department, Northern Illinois University, DeKalb, IL 60115, USA}
\author{D. E. Bugaris}
\affiliation{Materials Science Division, Argonne National Laboratory, Argonne, IL 60439-4845, USA}
\author{S. Avci}
\affiliation{Materials Science Division, Argonne National Laboratory, Argonne, IL 60439-4845, USA}
\affiliation{Department of Materials Science and Engineering, Afyon Kocatepe University, 03200 Afyon, Turkey}
\author{D.Y. Chung}
\author{H. Claus}
\affiliation{Materials Science Division, Argonne National Laboratory, Argonne, IL 60439-4845, USA}
\author{C. dela Cruz}
\affiliation{Quantum Condensed Matter Division, Oak Ridge National Laboratory, Oak Ridge, TN 37831, USA}
\author{M. G. Kanatzidis}
\affiliation{Materials Science Division, Argonne National Laboratory, Argonne, IL 60439-4845, USA}
\affiliation{Department of Chemistry, Northwestern University, Evanston, IL 60208-3113, USA}
\author{S. Rosenkranz}
\author{R. Osborn}
\affiliation{Materials Science Division, Argonne National Laboratory, Argonne, IL 60439-4845, USA}
\author{O. Chmaissem}
\affiliation{Materials Science Division, Argonne National Laboratory, Argonne, IL 60439-4845, USA}
\affiliation{Physics Department, Northern Illinois University, DeKalb, IL 60115, USA}

\date{\today}
\begin{abstract}
We present neutron diffraction analysis of  BaFe$_2$(As$_{1-x}$P$_x$)$_2$ over a wide temperature (10 to 300 K) and compositional ($0.11 \leq x \leq 0.79$) range, including the normal state, the magnetically ordered state, and the superconducting state.  The paramagnetic to spin-density wave and orthorhombic to tetragonal transitions are first order and coincident within the sensitivity of our measurements ($\sim 0.5$ K). Extrapolation of the orthorhombic order parameter down to zero suggests that structural quantum criticality cannot exist at compositions higher than $x = 0.28$, which is much lower than values determined using other methods, but in good agreement with our observations of the actual phase stability range. The onset of spin-density wave order shows a stronger structural anomaly than the charge-doped system in the form of an enhancement of the $c/a$ ratio below the transition.
\end{abstract}



\maketitle

\section{Introduction\label{intro}}

One topic of interest in the so-called 122 iron-based superconductors (compounds of the formula $A$Fe$_2$As$_2$, where $A$ is an alkaline earth) is to understand the underlying physics of the spin-density wave (SDW) ground state and the many different avenues that suppress it in favor of superconductivity.\cite{Johnston_puzzle_2010, Stewart_Superconductivity_2011} The charge-doped regimes have already undergone a lot of study, though the underlying chemistry of introducing charge can complicate the interpretation. \cite{Allan_Anisotropic_2013} Recently, isovalent substitution of P for As has received scrutiny\cite{Chaparro_Doping_2012, Fang_Doping-_2011, Moon_Infrared_2012, Nakajima_Crossover_2013} because the electronically clean substitution may make possible the detection of a potential quantum critical point (QCP) that arises from the suppression of the AFM phase transition down to 0 K somewhere near the composition of maximum $T_{\rm{c}}$ ($x\approx0.33$). \cite{Iye_gradual_2012, Walmsley_Quasiparticle_2013, van_der_beek_Quasiparticle_2010} Also, indications of nodal superconductivity suggest that the nature of the superconducting (SC) phase is different in this regime than the others.\cite{zhang_nodal_2012, Qiu_Robust_2012} This system shows the major features familiar to the other 122 substitution regimes: suppression of SDW, superconductivity that emerges before complete suppression of SDW order, and a superconducting dome that reaches a maximum not far after complete suppression of SDW order. Given that a primary chemical difference between As and P is that the latter is smaller, it has been postulated that the phase diagram is analogous to the pressure/temperature phase diagram. Previous researchers noted that P doping strains the lattice anisotropically, and it seems to mimic the effects of uniaxial pressure in the basal plane. \cite{Nakashima_Gigantic_2010, Rotter_different_2010, Zinth_Interplay_2012} In particular, they compared the Fe-As bond geometry and found a correlation with superconductivity between mechanical pressure and phosphorus doping.  

The onset temperature of orthorhombicity and magnetism (at temperatures $T_{\rm{s}}$ and $T_{\rm{N}}$, respectively) still remain contentious; some researchers have reported separate values differing by more than 10 K\cite{Iye_Microscopic_2012, Kasahara_Electronic_2012, Nakai_Normal-state_2013} while others report that they are the same.\cite{Kuo_Magnetoelastically_2012, Bohmer_thermodynamic_2012} Also, while the existing comparisons to mechanical pressure are appealing, they do not factor in how the Fe-P interaction contributes to the overall picture, except to say that it appears to enhance the features in the phase diagram more than expected by looking solely at the Fe-As bond. Given the sensitivity of the family to subtle changes in the bond geometry, the fact that there are two fundamentally different types of bonds randomly distributed throughout the material\textemdash unlike the mechanically strained system where the (Fe$_2$As$_2$) layer remains intact, structurally\textemdash suggests that in order to truly understand the system, one must develop a framework for understanding the Fe-P interaction as a function of doping.

As a step towards clarifying the nature of the transitions in this family, high-resolution neutron powder diffraction (NPD) was employed to monitor the SDW transition as a function of temperature and composition.  In this technique the nuclear and magnetic structures are measured simultaneously, which provides an internally consistent avenue for determining $T_{\rm{N}}$ and $T_{\rm{s}}$. We observe that the magnetic and structural transitions are coincident and weakly first order, and that the orthorhombic order parameter drops linearly as a function of composition, giving a zero-temperature intercept below $x = 0.28$.

\section{Experimental}

Two different methods were employed to synthesize polycrystalline samples of BaFe$_2$(As$_{1-x}$P$_x$)$_2$.  The samples with compositions of $x =  0.24$ and 0.25 were prepared by direct combination of the elements.  Stoichiometric ratios of the elements were slowly heated to 850 \textdegree C, whereupon they were soaked for 24 h, before turning off the furnace and allowing them to cool to room temperature.  The partially reacted materials were then homogenized by grinding with a mortar and pestle to a fine gray powder.  The material was then heated in a furnace at temperatures of 1000 \textdegree C (48 h), 1050 \textdegree C (48 h), 1100 \textdegree C (72 h), 1120 \textdegree C (120 h), and 1125 \textdegree C (twice, for 120 h and 24 h, respectively), with intermittent grinding between heat treatments.  For the samples with nominal compositions of $x$ = 0.115, 0.19, 0.205, 0.29, 0.31, 0.35, 0.37, 0.60, and 0.79, the phase-pure ternary compounds BaFe$_2$As$_2$ and BaFe$_2$P$_2$ were first prepared from the elements by heating at temperatures of 750 \textdegree C (72 h), 1080 \textdegree C (48 h), and 1100 \textdegree C (60 h).  Stoichiometric ratios of BaFe$_2$As$_2$ and BaFe$_2$P$_2$ were ground together with a mortar and pestle, and heated in a furnace at a temperature of 1120 \textdegree C (three or four cycles, with durations of 80-96 h), with intermittent grinding between heat treatments.

Handling of all starting materials was performed in an Ar-filled glovebox. For furnace heating cycles below 1000 \textdegree C, the reactants were contained in alumina crucibles that were sealed in quartz tubes under vacuum. In order to heat at temperatures of 1000 \textdegree C or above, the reactants in the alumina crucibles were first sealed in Nb tubes that were welded shut under Ar, and then subsequently sealed in quartz tubes under vacuum. The sealed Nb tubes were necessary to prevent non-stoichiometry by volatilization of As or P and reaction with the quartz at high temperatures.

Initial characterization of the dark gray powders was conducted by powder x-ray diffraction using a Panalytical X'pert Pro diffractometer with an iron filtered Cu-Kα source. Magnetization measurements were conducted at 0.1 Oe on a home-built SQUID magnetometer.

Time-of-flight NPD experiments were performed on the POWGEN beamline at the Spallation Neutron Source at Oak Ridge National Laboratory (ORNL).  The temperature dependent scans were collected on warming. For the improved statistics on the $x$ = 0.19 sample, temperature dependent intensity around the (103) reflections was measured on beamline HB-1A at the High-Flu x Isotope Reactor at ORNL. The POWGEN data was fit using the Rietveld method in GSAS\cite{GSAS} and EXPGUI\cite{expgui}.  Peaks shapes were modeled using back-to-back exponentials convoluted with a pseudo-Voigt and employing microstrain broadening. \cite{stephens_1999}

\section{Results}

\subsection{Superconductivity and stoichiometry}
\begin{figure}
\includegraphics[width=8.5cm]{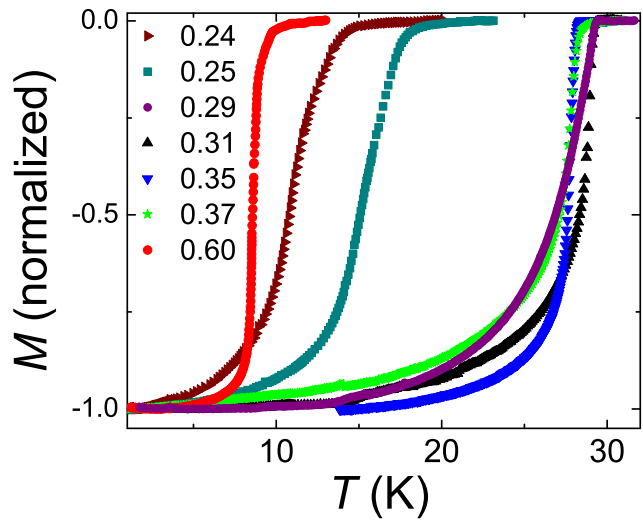}
\caption{Normalized magnetization measurements on representative samples. Superconducting volume fractions are estimated using the magnitude of the diamagnetic response and are $\geq 80\%$ for all samples. \label{magnetization}}
 \end{figure}

The neutron diffraction experiments determined precise lattice parameters over a wide range of temperatures and compositions and were sensitive to magnetic moments larger than $\sim0.3~\mu_B$ in these materials. The prepared compounds exhibited very sharp reflections indicative of the good quality of the samples. BaFe$_2$(As$_{1-x}$P$_x$)$_2$ follows Vegard's Law,\cite{Nakajima_Growth_2012, Kasahara_Evolution_2010} so the determined cell volumes were used to find the actual stoichiometry. Using either a simple interpolation of the known endpoint volumes\cite{Avci_Phase_2012, Mewis_Ternary_1980} or a linear fit of all of our determined volumes based on nominal compositions give actual compositions that agree with each other within 0.003. We report the actual composition to the nearest 5 on the third decimal place. Some powders showed Fe$_2$P as a minority impurity phase ($\leq 5\%$), though only a few samples differed more than 0.01 from nominal. These same samples tended to show heightened impurity content. 

Sample quality was also confirmed by checking the superconductivity properties of the appropriate samples. Superconductivity was not observed for $x = 0.115$, 0.19, 0.205, and 0.79, whereas sharp superconducting transitions were observed for $x = 0.24$, 0.25, 0.29, 0.3, 0.31, 0.35, 0.37, and 0.60 (Figure \ref{magnetization}). Using the criterion where $T_{\rm{c}}$ is defined as the point where 10\% of the maximum diamagnetic signal is achieved gives transition temperatures of 13.1, 17.2, 29, 29.2, 28, 28.1, and 9.07 K, respectively.  For $x=0.24$ and 0.25 the transitions are slightly broader, though still much sharper than most previous reports for this sample range, indicative of good sample quality.\cite{Iye_gradual_2012, Iye_Microscopic_2012, Bohmer_thermodynamic_2012, Nakajima_Growth_2012} It was shown previously that the anomalous broadening of transitions in this regime originates from enhanced coupling between strain and the superconducting order parameter\cite{Kuo_Magnetoelastically_2012}, though most likely the rapid variation of $T_{\rm{c}}$ as a function of composition also allows small compositional inhomogeneities to be observed as a broadened transition.  Nevertheless, even for such closely spaced compositions, the curves for $x = 0.24$ and 0.25 are sufficiently well spaced to indicate that they are of distinguishable composition, and that the samples have a smaller than usual concentration of internal strain inducing defects. 

\subsection{Structure\label{ResStruct}}

\subsubsection{Crystal structure}

 \begin{figure}
 \includegraphics[width=8.5cm]{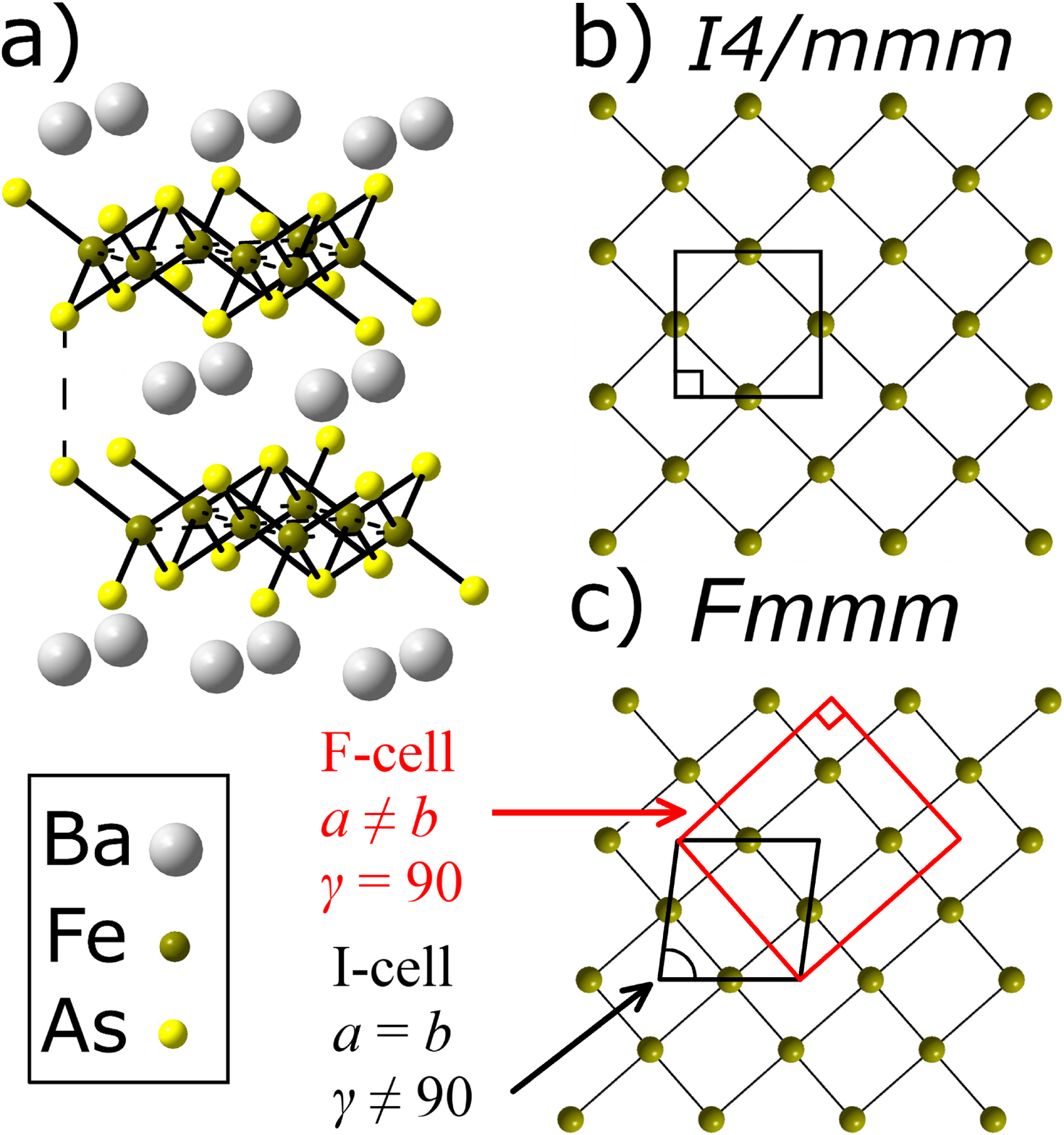}
 \caption{a) Depiction of the crystal structure of BaFe$_2$As$_2$.  b) Basal plane lattice vectors and Fe sublattice of the $I4/mmm$ cell viewed along the (001) direction. c) $Fmmm$ cell in the same view.  The symmetry breaks when $\gamma$ diverges from 90\textdegree, though primitive $a$ and $b$ axes remain the same which gives a conventional orthorhombic cell. The magnitude of the distortion is exaggerated for clarity.  \label{unitcell}}
 \end{figure}

The crystal structure for these phases is already well known and served as the basis for the Rietveld refinements. For the ensuing discussion, the most relevant details of the crystal structures are summarized here. The crystal structure of the body-centered tetragonal phase shown in Figure \ref{unitcell} is the ThCr$_2$Si$_2$-type structure (space group $I4/mmm$). The (Fe$_2$As$_2$) layer is made up of edge-sharing FeAs$_4$ tetrahedra ($d_{Fe-As}\approx2.4$ \AA) and the Ba atoms occupy nearly cubic cages between the layers. The square lattice of iron atoms has a nearest-neighbor spacing of $d_{Fe-Fe} = a / \sqrt{2}\approx 2.75$ \r{A}, meaning that there is significant direct orbital overlap between metals.  The interlayer interactions are significantly weaker; the nearest spacing is non-bonding ($d_{As-As} \approx 3.8$ \r{A}), meaning that it is the more ionic Ba-As bond ($d_{Ba-As} \approx 3.3$ \AA)that holds the layers together. 

The structural transition lowers the symmetry from tetragonal to orthorhombic symmetry (space group $Fmmm$).  The distortion leaves the magnitudes of the tetragonal $a$ and $b$ axes equivalent, but the angle between them, $\gamma$, diverges slightly from 90\textdegree. This reduces the order of the principal rotational symmetry axis from $C_4$ to $C_2$. Though translational symmetry is preserved, the conventional cell is twice as large $\boldsymbol{a_{orth}} = \boldsymbol{a_{tet}} -\boldsymbol{b_{tet}}$; $\boldsymbol{b_{orth}} = \boldsymbol{a_{tet}} + \boldsymbol{b_{tet}}$, which yields the $F$-centered cell (Figure \ref{unitcell}c). This setting makes it clear that the underlying distortion is the formation of orthorhombic stripes in the Fe sublattice (i.e., the bonds lengthen in one direction and shorten in the other) suggesting that metal-metal interactions are the driving force in the transition. Magnetically, the moments are ferromagnetically coupled in the $b_{\rm{orth}}$ (short) direction and antiferromagnetically coupled in the $a_{\rm{orth}}$ (long) direction\textemdash which is also the axis that the moments are oriented along. Aside from the creation of two inequivalent Fe-Fe bonds from a single one above $T_{\rm{s}}$, the only other symmetry allowed change is that one of the two unique tetrahedral bond angles, $\alpha_2$, splits into $\alpha_{2}^\prime$ and $\alpha_{2}^{\prime\prime}$.

The only refinable atomic position in either space group is the $z$ parameter on the $X$ site, ($X$ = As, P). In BaFe$_2$As$_2$ this is fully occupied by As, but the addition of P as a dopant introduces disorder on this site. Typical Fe-As and Fe-P bond distances are quite different (2.40 vs 2.26 \r{A} based on the end members), so one might expect each anion to contribute to a different average $z$ parameter, giving two distinct sites.  Indeed, a previous single-crystal x-ray diffraction study has reported separate $z$ values for the As and P sites at a few compositions.\cite{Rotter_different_2010}

Unfortunately, using only neutron powder data both sites could not be reliably refined separately, and instead the As and P sites were constrained to be equivalent. The neutron cross-sections of P and As are much closer than their x-ray scattering factors, so the averaged position cannot be directly compared to an averaged site obtained using x-rays. The refined $z$ parameters are reported in Table \ref{tablelattice}, though the value is not particularly meaningful when taken naively, because the actual values for As and P ought to be quite different.  Likewise, the averaged-site $X$-Fe-$X$ bond angles are not reported here as a function of composition because they are not physically meaningful.

\subsubsection{Refined lattice parameters}

  \begin{figure}
 \includegraphics[width=8.8cm]{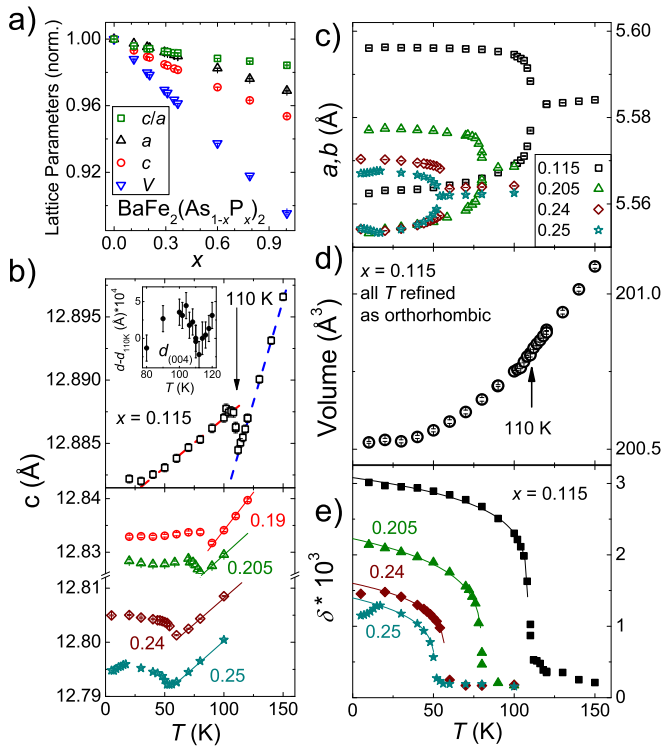}
 \caption{(a): Normalized lattice parameters as a function of composition at 300 K. The BaFe$_2$As$_2$ point is taken from Ref. \cite{Avci_Phase_2012} and the BaFe$_2$P$_2$ point is taken from Ref \cite{Mewis_Ternary_1980}. (b) $T$ dependence of $c$-axis of $x=0.115$ (top panel) and $x=$ 0.19, 0.205, 0.24, and 0.25 (bottom panel). The  inset shows the temperature dependence of (004) reflection peak position.  (c) Dependence of $a_{\rm{orth}}$ and $b_{\rm{orth}}$ axes as a function of temperature. For the tetragonal data points, $a$ is scaled by a factor of $\sqrt{2}$ in order to compare equivalent lattice vectors. (d) Volume of $x = 0.115$ as a function of temperature. In order to avoid artificial discontinuities when changing from one model to the other, only orthorhombic refinements are used for all temperatures. The volume is given in the body-centered cell by scaling by 0.5. (e) The orthorhombic order parameter is shown for $x = 0.115$, 0.205, 0.24, and 0.25.  \label{lattice}}
 \end{figure}

Refined crystal structure parameters determined at 10 and 300 K are reported in Table \ref{tablelattice}, whilch illustrates the Vegard's Law behavior across the entire phase diagram (Figure \ref{lattice}a). As mentioned above, the compositions in this study are defined based on the 300 K Vegard's law trend, though the trend itself is the same regardless of corrections.  Note that $c$ is more sensitive to P substitution than $a$, which gives the $c/a$ ratio a slightly non-linear shape. The temperature dependence of the lattice parameters is shown in Figures \ref{lattice}b-d.  The $a$- and $b$-axes split at $T_{\rm{s}}$, and the $c$-axis shows some negative thermal expansion just below this temperature. At higher doping, the negative expansion of $c$ continues to the lowest measured $T$, giving an enhanced value of approximately 0.01 \r{A} compared to usual normal state behavior of $c$ at low $T$ (i.e. from samples that do not exhibit orthorhombicity).  The nature of the transition can also be elucidated by observing the slope of $c(T)$. Above $T_{\rm{s}}$ it is significantly larger than below (Figure \ref{lattice}b), and at $T_{\rm{s}}$ there is a discontinuity in going from one phase to the other.  The refined $c$ parameters in this region are shown in Figure \ref{lattice}b. In order to rule out that this discontinuity is an artifact of the change in model from $I4/mmm$ to $Fmmm$, the position of the (004) reflection is plotted in the inset, which shows that there is a sharp reversal in slope at $T_{\rm{s}}$ regardless of model used.  For $x = 0.115$ the fine steps span the whole negative slope region, which is $\sim$ 10 K across. The volume also shows a weak feature around $T_{\rm{s}}$ (Figure \ref{lattice}d).

Significantly below $T_{\rm{s}}$, orthorhombicity is obvious due to the splitting in peaks with $h,k \neq 0$. When approaching the transition the splitting continuously decreases, which means that the precise point of the transition to $I4/mmm$ cannot be determined directly.  Instead it is extrapolated by determining the magnitude of the orthorhombic distortion as a function of temperature. The orthorhombic order parameter $\delta=\frac{(a-b)}{(a+b)}$ is calculated from fits to an orthorhombic model at each temperature and plotted in Figure \ref{lattice}e. It is clear that well above $T_{\rm{s}}$ that the refined $\delta$ is nearly constant, which is a reflection of the peak width limit for this experiment.  At low temperature $\delta$ is saturated, but near $T_{\rm{s}}$ it drops rapidly, a behavior that can be parameterized using the power law $\delta(T)=A_{\rm{s}}(T_s-T)^{\beta_{\rm{s}}}/T_s$, which gives a precise estimation of $T_{\rm{s}}$. This is tabulated in Table \ref{tabletransitions} for fits using only data points below $T_{\rm{s}}$  and above $T_{\rm{c}}$ (for $x=0.24$ and 0.25). Fits agreed well with the data (0.985 $\leq R^2 \leq$ 0.9998)

Note that there are a few points above $T_{\rm{s}}$ that exhibit enhanced $\delta$ compared to the baseline which are a result of anisotropic peak broadening. This is not surprising, because magnetostructural transitions are often accompanied by such broadening which is known to be a result of non-cooperative static fluctuations that locally break the symmetry.\cite{Giot_Magnetoelastic_2007, Khalyavin_Spin-ordering_2011} Kasahara et al. observed similar broadening in this family within the same temperature region using synchrotron x-rays, \cite{Kasahara_Electronic_2012} which they ascribed to a more complex electronic-nematic phase where the electronic structure is ordered but the nuclear structure is not. They further suggested that the ``true'' phase transition is where this broadening extrapolates to the baseline peak width.  Neutrons are only sensitive to the positions of the nuclei, so compared to their data it appears that the nuclear and electronic anisotropy are approximately equivalent. This means that the data reported here does not distinguish between a standard magnetostructural microstrain model and the more complex electron-nematic model proposed in Ref. \cite{Kasahara_Electronic_2012}, though based on our structural data it does appear that the true phase transition really is at $T_{\rm{s}}$ and not at some higher $T^{\rm{*}}$ (See Section \ref{structureresults} for more details).

\begin{table*}
 \caption{Lattice parameters of BaFe$_2$(As$_{1-x}$P$_x$)$_2$ at 10 and 300 K refined from NPD data (POWGEN) using determined $x$ values.  \label{tablelattice}}
 \begin{ruledtabular}
 \begin{tabular}{c c c c c c}
\emph{x} & \emph{a} (\r{A}) & \emph{b} (\r{A}) & \emph{c} (\r{A}) & \emph{V} (\r{A}$^3$) & $z_{As}$ \\
\hline
\multicolumn{6}{c}{10 K} \\
\hline
0.115 & 5.59609(5) & 5.56245(5) & 12.88371(15) & 200.522(3) & 0.35336(7) \\
0.190 & 5.58034(3) & 5.55570(5) & 12.83531(11) & 198.685(5) & 0.35277(6) \\
0.205 & 5.57761(3) & 5.55369(3) & 12.82837(8)   & 198.736(3) & 0.35265(4) \\
0.24   & 5.57037(3) & 5.55420(3) & 12.80496(9)   & 198.086(2) & 0.35231(5) \\
0.25   & 5.56733(4) & 5.55393(4) & 12.79542(12) & 197.821(3) & 0.35213(6) \\
0.290 & 5.55378(2) & 5.55378(2) & 12.75240(12) & 196.670(2) & 0.35191(5) \\
0.310 & 5.55094(2) & 5.55094(2) & 12.75590(8)   & 196.523(1) & 0.35178(5) \\
0.350 & 5.54439(2) & 5.54439(2) & 12.73346(9)   & 195.715(2) & 0.35141(6) \\
0.370 & 5.54035(3) & 5.54035(3) & 12.72376(16) & 195.281(3) & 0.35113(10) \\
0.600 & 5.49977(2) & 5.49977(2) & 12.59785(9)   & 190.527(2) & 0.34946(7) \\
0.790 & 5.46401(4) & 5.46401(4) & 12.5054(2)     & 186.678(4) & 0.34775(10) \\
\hline
\multicolumn{6}{c}{300 K}  \\
\hline
0.115 & 3.952503(12) &-& 12.95775(12) & 202.430(2) & 0.35347(7)\\
0.190 & 3.943912(12) &-& 12.90799(12) & 200.777(2) & 0.35282(6)\\
0.205 & 3.941969(15) &-& 12.90210(10) & 200.487(2) & 0.35286(6)\\
0.290 & 3.931949(11) &-& 12.84854(8)   & 198.641(3) & 0.35205(5)\\
0.310 & 3.929868(12) &-& 12.83854(8)   & 198.277(2) & 0.35191(5)\\
0.350 & 3.925049(14) &-& 12.81491(9)   & 197.427(2) & 0.35161(6)\\
0.370 & 3.922218(19) &-& 12.80440(14) & 196.980(2) & 0.35140(9)\\
0.600 & 3.893640(16) &-& 12.66917(12) & 192.070(2) & 0.34917(4)\\
0.790 & 3.86863(4)	    &-& 12.56670(19) & 188.077(3) & 0.34749(11)\\

 \end{tabular}
 \end{ruledtabular}
 \end{table*}
\begin{table*}
 \caption{Refined magnetic moment and parameters to power-law fit to structural and magnetic order parameters. $M_{\rm{eff}}$ refers to the 10 K refined value. For $x$ = 0.190 the POWGEN data was not finely spaced enough to be reliably fit, so only the magnetic HB-1A data was fit to a power-law.\label{tabletransitions}}
 \begin{ruledtabular}
 \begin{tabular}{c c c c c c c c}
\emph{x} & $T_{\rm{s}}$ (K) & $\beta_{\rm{s}}$ & $A_{\rm{s}}$*10 & $M_{\rm{eff}}(\mu_{\rm{B}})$ & $T_{\rm{N}}$ (K) & $\beta_{\rm{N}}$ & $A_{\rm{N}}$ \\
\hline
0.115 &110.02(11) & 0.134(8)  & 1.82(6)    & 0.75(4) &110.4(8)& 0.077(14)&60(3)\\
0.190 &-                  & -              & -               & 0.51(4) & 82.8(10) & 0.14(3)    &23(3)\\
0.205 &80.03(7)     &0.180(10) & 0.81(3)    & 0.53(4) & 80.0(3) & 0.05(2)    &26(3)\\
0.24   &57.4(4)       &0.173(6)   & 0.456(7)  &  0.46(4)&-             & -               &-       \\
0.25   &51.13(7)     & 0.198(3)  &0.328(3)   & 0.43(5) &-             & -               &-       \\
 \end{tabular}
 \end{ruledtabular}
 \end{table*}

\subsection{Magnetism}
 \begin{figure}
 \includegraphics[width=9cm]{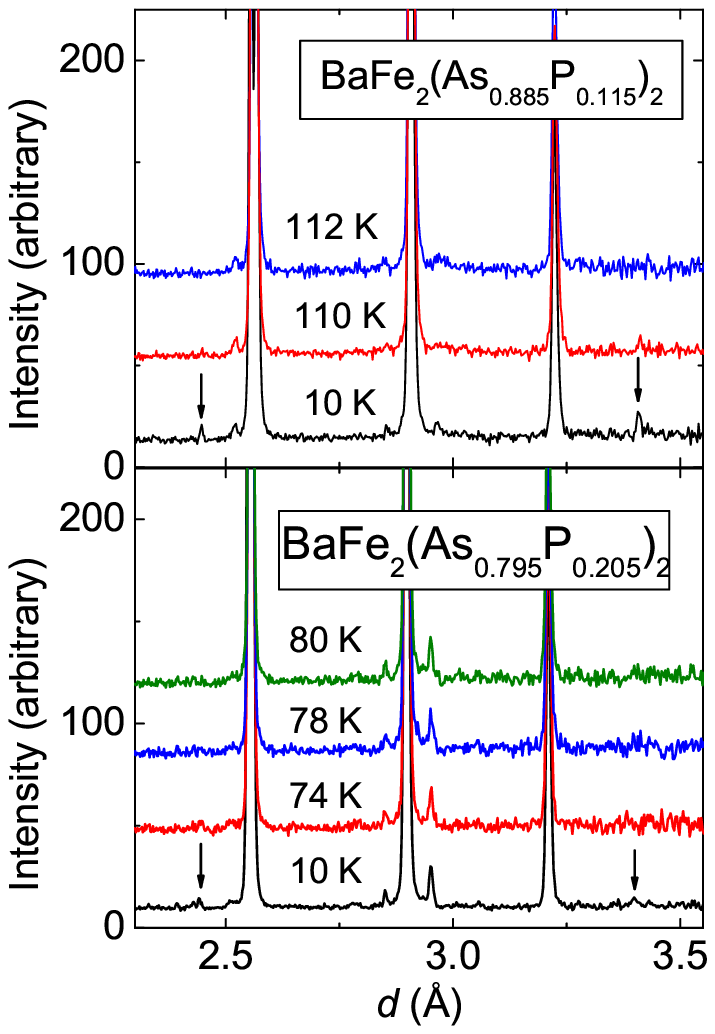}
 \caption{Histograms of the $x = 0.115$ (upper panel) and 0.205 (lower panel) POWGEN data at selected temperatures. Arrows point to the magnetic peaks in the 10 K scans.
\label{mag1}}
 \end{figure}
 
\begin{figure}
 \includegraphics[width=9cm]{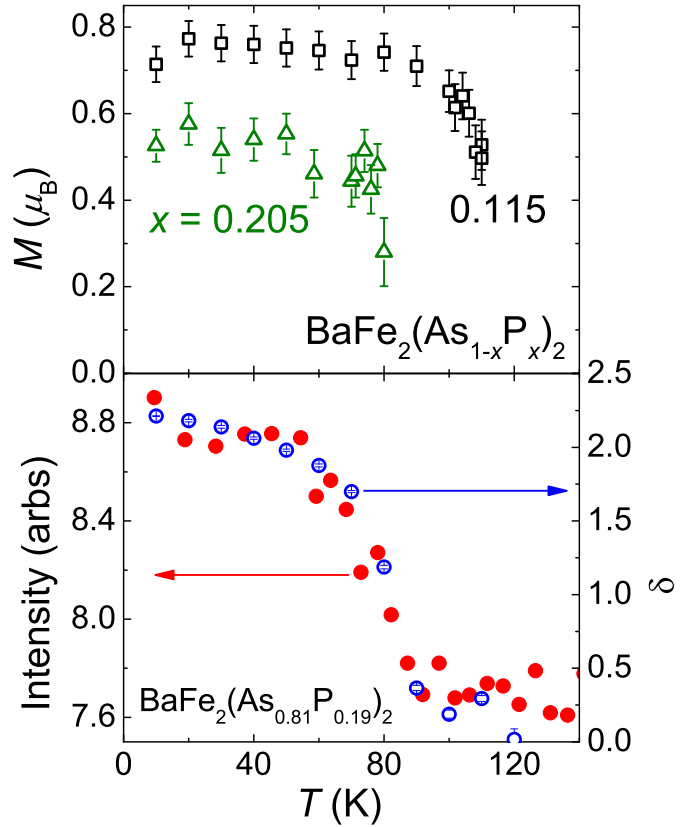}
 \caption{The refined magnetic moments for $x = 0.115$ and 0.205 from the Rietveld refinements are shown in the upper panel.  The lower panel shows an overlay of the integrated intensities of the (103) magnetic reflection from the HB-1A experiment (closed red circles) and the refined orthorhombic order parameter (open blue circles). (Color online).
\label{mag2}}
 \end{figure}

The orthorhombic model shows splitting of major reflections, but it does not introduce any new supercell reflections, so new reflections at  low temperature are used to identify the magnetic phase.  The magnetic phase was refined for these samples in the magnetic space group $F_Cmm^\prime m^\prime$ resulting in good agreement with the data.  Even at low $T$ the magnetic peaks are very weak, so approaching $T_{\rm{N}}$ it becomes increasingly difficult to observe the reflections (Figure \ref{mag1}). For $x$ = 0.115 and 0.205 samples enough statistics and temperature points were collected in order to both refine the moments (Figure \ref{mag2}, upper panel) and fit them to the same power law used in Section \ref{ResStruct}.    No magnetic peaks can be distinguished from the noise above $T = 110$ and 80 K, for $x$ = 0.115 and 0.205, respectively.  The 10 K refined magnetic moments and results of the power law fit are given in Table \ref{tabletransitions}.    For $x$ = 0.24 and 0.25 less material was available, and the magnetic order parameter had weakened by this point, so the magnetic reflections could only be seen in the long 10 K collection. They were too weak to refine during the temperature-dependent portion of the experiment, so the magnetic transitions were not determined for these compositions.

Data were collected on HB-1A in order to improve the intensity statistics on the (103) magnetic peak of our $x = 0.19$ sample, which gives a fitted transition of 82.8(10) K (the magnetic moment is still distinguishable from the noise at 82.2 K, but not at the next temperature increment at 87.5 K). The lower panel of Figure \ref{mag2} shows the integrated intensity under this reflection superimposed on the orthorhombic order parameter from the POWGEN refinements, which agree well.

\section{Discussion\label{discussion}}
\subsection{Magnetic and structural transitions\label{structureresults}}
\subsubsection{Phase Diagram}

\begin{figure}
\includegraphics[width=8.5cm]{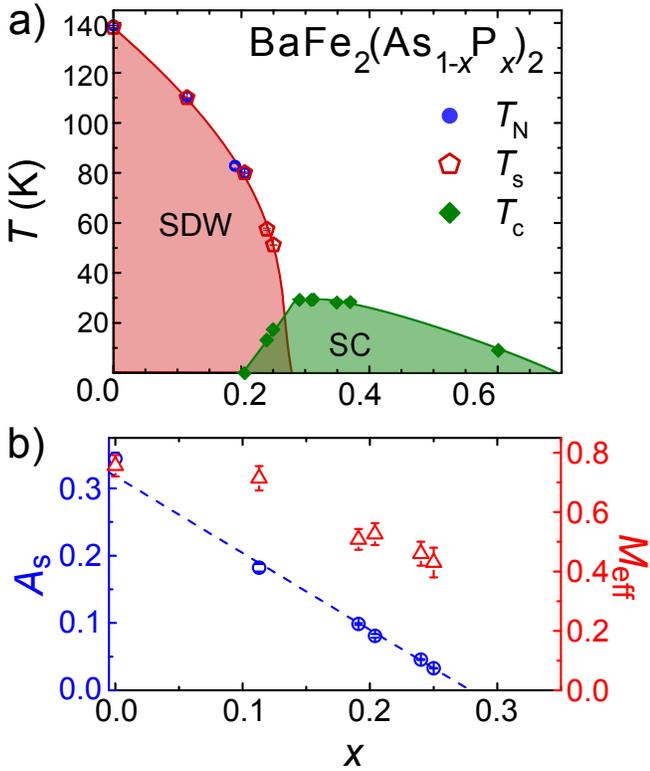}
\vspace{-5pt}
\caption{a) The phase diagram of BaFe$_2$(As$_{1-x}$P$_x$)$_2$. Phase lines are drawn to guide the eye. Error bars for both axes are contained within the symbols. b) Composition dependence of the orthorhombic order parameter prefactor, $A_{\rm{s}}$ (blue), and the 10 K refined magnetic moment (red).  \label{phasediagram}}
\end{figure}

Taking all of our data together allows for the construction of a phase diagram which is plotted in Figure \ref{phasediagram}a. The magnetic transition temperatures, $T_{\rm{N}}$, agree closely with the respective $T_{\rm{s}}$'s. The high-resolution allows for more confidence in the orthorhombic order fits than the magnetic fits; still the error in fitted $T_{\rm{N}}$ is $<1$ K.  The order of the power law, $\beta_{\rm{N}}$, is less well defined due to the weak intensity of the reflections, though it appears small. This signifies that magnetic and structural transitions are the same for those measured here ( $x \leq 0.205$). The compositional dependence of $T_{\rm{s}}$ and $T_{\rm{N}}$ shows a nearly linear decrease as a function of composition up to $\sim0.2$, after which the suppression is enhanced further.  Between $x = 0.25$ and 0.29 the dying off of SDW order must become much steeper in order for orthorhombic order to be completely suppressed in this range.

The actual point of complete suppression of SDW order can be more precisely determined by looking at the rate of suppression of the relevant order parameter. Since the value of $\delta$ changes as a function of temperature, and competes with superconductivity in some samples, a temperature-independent form is needed to compare the strength of the orthorhombic distortion across different compositions. One estimation that can be used is the prefactor of the power-law fit, $A_{\rm{s}}$, which is linear in the observed range (Figure \ref{phasediagram}b). This extrapolates to zero at $x = 0.279(12)$ ($R^2 = 0.995$), implying that there is no orthorhombic state at all above this composition. Additionally, this is based solely on data from the normal, SDW state, and for $x = 0.24$ and 0.25 there is a suppression of $\delta$ below $T_{\rm{c}}$ that is not taken into account here. This is typically understood to give rise to a reversal in sign of the SDW phase line below the SC phase line\cite{Nandi_Anomalous_2010}, which means that taking this into account will push critical composition to still lower values.  Scaling the compositional range to that observed in Ba(Fe$_{1-x}$Co$_x$)$_2$As$_2$\cite{Nandi_Anomalous_2010} gives a shift of approximately 0.01, yielding a projected critical composition of $x_{\rm{c}}\approx 0.27$.  As for the magnetic order parameter, power laws fits to the magnetic moments are not possible for all compositions, so the 10 K magnetic moments were used instead. The behavior is clearly non-linear; instead it shows a power-law behavior similar to the hole-doped systems. The statistics are not sufficient for fitting, though it should be noted that when the observed peak is near the detection limit the refinement tends to overestimate the effective moment. This means that for $x=$ 0.24 and 0.25, the actual moment may be smaller than reported here. 

Previous studies have found evidence of quantum critical fluctuations that they interpreted as the existence of a QCP arising from suppression of SDW order that coincides with the maximum $T_{\rm{c}}$ at or around $x = 0.33$\cite{Kasahara_Evolution_2010, Iye_gradual_2012, Walmsley_Quasiparticle_2013}. On the other hand, we found that the critical composition is likely between $x=0.27$ and 0.28 and we also see a consistent trend of increasing $T_{\rm{c}}$ for our $x = 0.29$, 0.31, and 0.315 (not shown) samples and then a decrease again at $x=0.35$. Moreover, none of these samples exhibit orthorhombic or magnetic order that we can measure, which is internally consistent with our projected $x_{\rm{c}}$. This indicates that the composition of our maximum $T_{\rm{c}}$ is in close agreement to these previous studies, and it casts doubt on the idea that there is a QCP that occurs at the point of maximum $T_{\rm{c}}$. That is, unless the observations of quantum critical fluctuations are entirely spurious, the only way to lift this inconsistency is to allow the QCP to be at a significantly lower composition than the point of maximal $T_{\rm{c}}$. Another possible explanation is that the critical fluctuations above $x = 0.3$ are real, but that they represent a different type of order than is measured in this experiment. For example, it could represent magnetic order in the absence of orthorhombic order of the type that we measured recently in the Ba$_{1-x}$Na$_x$Fe$_2$As$_2$ system.\cite{avci_magnetically_2014} Our projected value for the complete suppression of orthorhombic order, $x_{\rm{c}}$, is very robust, because it comes from the scale of the power-law fits to the orthorhombic order parameters, which have excellent agreement to the data. Likewise, high-resolution neutron diffraction gives very precise and reproducible orthorhombic order parameters, which in general are less instrument-dependent than raw lattice parameters.

 \begin{figure}
 \includegraphics[width=9.4cm]{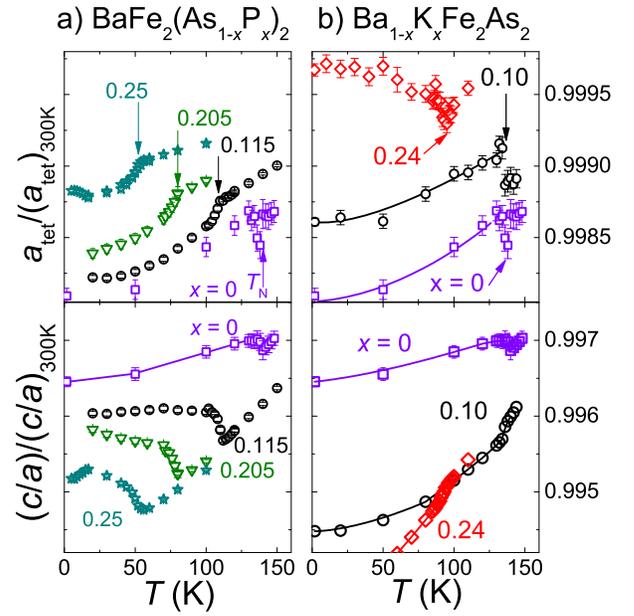}
 \caption{Normalized lattice parameters as a function of temperature of a) BaFe$_2$(As$_{1-x}$P$_x$)$_2$ and b) Ba$_{1-x}$K$_x$Fe$_2$As$_2$ ($x$ = 0, 0.1, 0.15, and 0.24). The upper panels show the $a_{\rm{tet}}$ normalized to each composition's room temperature value, and the lower panels show $c/a$ ratios normalized in the same manner. The data points for $x=0$ and the K-substituted compounds is taken from Ref \cite{Avci_Phase_2012}. The lines are guides to the eye. \label{lattnorm}}
 \end{figure}

\subsubsection{First-order phase transition}

The discontinuities in the individual lattice parameters are unambiguous, and coincide very well with the $T_{\rm{s}}$'s determined using the power law fit to $\delta$. As discussed in Section \ref{ResStruct}, the c-axis has a clear discontinuity at $T_{\rm{s}}$. The reason that there is not a similar change in volume can be seen when the body-centered $a_{\rm{tet}}$ is plotted.  As $T$ is decreased there is a sudden increase in the slope of $a_{\rm{tet}}(T)$ at $T=T_{\rm{s}}$ that eventually relaxes to a more typical value by 25 K. The feature is not as drastic as the increase in $c$, but the discontinuous slope at $T=T_{\rm{s}}$ is clear. This drop compensates the increase in $c$ which is why the feature is weak and somewhat ambiguous, unlike previous high resolution diffraction on the hole-doped materials Ba$_{1-x}A_x$Fe$_2$As$_2$ (\emph{A} = Na, K), which showed a discontinuous drop in volume in the transformation from orthorhombic to tetragonal symmetry \cite{Avci_Phase_2012, Avci_Structural_2013}.

Given that the changes between basal plane and interlayer spacings at the transition are compensatory, their combined effect is best seen by plotting the $c/a$ ratio (using $a=a_{\rm{tet}}$) in Figure \ref{lattnorm}a. In the normal state $c/a$ rises monotonically up to room temperature, but at $T_{\rm{s}}$ there is a discontinuity to a higher value in the orthorhombic state.   Below $T_{\rm{c}}$, the $c/a$ ratio is seen to be suppressed towards the $T>T_{\rm{s}}$ phase line in a fashion similar to what is seen for $\delta$ and $M_{\rm{eff}}$.  This confirms that these more subtle distortions in the structure are coupled with the orthorhombic and magnetic order parameters, and are also partially suppressed by the superconducting state in the same way. Note that the discontinuity at $T_{\rm{s}}$ is also seen in the parent compound as a weak trough. The hole-doped compounds are shown in Figure \ref{lattnorm}b for comparison.  Interestingly, even though there is evidence of some strange behavior in the slope of $a_{\rm{tet}}$ at low $T$, the feature is barely noticeable in the $c/a$ plot and it is of the opposite sign.

 The discontinuities in the lattice parameters, visible both individually and in the $c/a$ ratio, together with the exponents from the power-law fits to the structural and magnetic order parameter, can be explained as a first-order phase transition where the high temperature (ht) paramagnetic tetragonal and low temperature (lt) antiferromagnetic orthorhombic phases are in coexistence.  Based on the extrapolated slopes from the ht and lt regimes, each phase has a slightly different $c$ axis in this area, but they are so close that no splitting is observed in the actual peak, only a slight broadening. The difference is too small to refine both components, so instead the refined lattice parameter arises from a convolution of the relative intensities of the overlapping reflections. We estimate that this negative slope regime corresponds roughly to the same temperature regime as the negative slope in resistivity observed by Iye et al.\cite{Iye_Microscopic_2012} and Nakajima et al.\cite{Nakajima_Growth_2012}, implying that both cases show the discontinuous transformation from the lt- to the ht-phase, though presumably heating/cooling schedules change the onsets somewhat. The $\beta_{\rm{s}}$ exponents determined from the power-law fits to the orthorhombic order are anomalously low, which is also consistent with a weakly first-order phase transition.

An alternate interpretation of this data, as a second-order phase transition, is not as well supported by the data.  This model requires the lattice parameters to vary continuously across the phase transition, meaning that the negative slope regime in $c(T)$ represents the actual behavior of the lattice parameter instead of the transfer of spectral weight from a ht- to lt-phase, each with different lattice parameters. The negative c-axis slope is uncorrelated with the progression of the orthorhombic order parameter, which suggests that the c-axis is not co-evolving with the magnitude of orthorhombic order. Also the crossover region would be expected to have a smoother shape to it characteristic of a continuous first derivative.  

\subsection{Comparison to other 122 systems\label{compare}}

The most intuitive analog to P-doping is to compare it to mechanical pressure.  It has already been shown that while the chemical pressure scenario does not work when comparing only cell volumes,\cite{Rotter_different_2010} anisotropic changes in the Fe$_2X_2$ basal plane correlate with similar changes from uniaxial pressure,\cite{Zinth_Interplay_2012} implying that the bond geometry is important to the properties.  It was noted that the Fe-As bond length in particular correlates to the superconducting dome magnitude and range under both the mechanical and chemical pressure regimes.  However, they did not include the effects of the Fe-P bond in their analysis, which should also have a significant effect on the properties.

As for other chemical doping regimes, hole-doping on the $A$-site\textemdash e.g. Ba$_{1-x}A_x$Fe$_2$As$_2$ ($A$ = Na,K)\cite{rotter_superconductivity_2008-1, rotter_superconductivity_2008-2, Cortes-Gil_indifference_2010}\textemdash has been found to suppress $T_{\rm{N}}$, the magnetic order parameter ($M_{eff}$), and the orthorhombic ordering parameter in a similar, power-law like fashion. The orthorhombic and magnetic transitions remain coincident over the entire range of the phase and the combined phase transition is weakly first-order.  On the other hand, Co (electron) doping on the Fe-site  suppresses both types of ordering nearly monotonically, and eventually separates the structural and magnetic transitions at a tricritical point\cite{pajerowski} into two, second-order transitions which eventually differ by over 10 K.\cite{Ni_Effects_2008, Nandi_Anomalous_2010}

For the BaFe$_2$(As$_{1-x}$P$_x$)$_2$ system a combination of these features is seen.  The composition dependence of the SDW is more akin to the electron-doped regime where it drops somewhat faster than linearly, but not obviously by a power-law.  The coincident, weakly first-order transitions are reminiscent of the hol- doped compounds, but unlike those materials there is not an obvious correlation between the compositional dependence of the magnetic and orthorhombic order parameters.  The overlap between the SDW and SC phases is perhaps smaller than it is in either of the charge-doped regimes, instead showing a rather sharp increase in the SC transition, along with a sudden disappearance of the SDW phase. The linear decrease of the magnitude of the orthorhombic order, $A_{\rm{s}}$, is remarkable since to our knowledge it is not seen in any other doping regime, and it implies that the phosphorus substitution is very efficient at disrupting structural order.

Likewise, the lattice anomalies observed at $T_{\rm{s}}$ are another unique feature: in the hole-doped system a much more ambiguous effect is seen, and it is of opposite sign and weakens substantially as a function of composition.  As can be seen in Figure \ref{lattnorm} the same weak trough in $a_{\rm{tet}}$ that is seen in the parent compound is seen in the Ba$_{1-x}$K$_x$Fe$_2$As$_2$ samples ($x$ = 0.1 and 0.24). Nearly identical trends are seen for the Ba$_{1-x}$Na$_x$Fe$_2$As$_2$ phases based on data from Ref \cite{Avci_Structural_2013}, so they are not shown. As for the $c/a$ ratio, it is less analogous to the parent compound.  In the parent compound there is a slight enhancement of $c/a$ upon cooling through the transition and then a decrease in slope, while in the hole-doped systems $c/a$ is slightly suppressed.  Unfortunately we have not been able to find previous reports on electron-doped materials that explicitly define $c/a$ the way we have here, so we cannot directly compare.  However, an extraction of the data plotted in Ref \cite{Zinth_Interplay_2012} indicates that even when phosphorus and cobalt are co-substituted, $a_{\rm{tet}}$ at $T_{\rm{s}}$ behaves similarly to how it does in the hole-doped system.

There are two primary interactions in these materials: Fe-As(P) and Fe-Fe bonds. Since it is ostensibly the nearest neighbor (nn) Fe-Fe interaction that drives the structural transition (by separating the square Fe net into stripes along the short axis), it is likely this explains the decrease in $a_{\rm{tet}}$. That is, there is an overall increase in the Fe-Fe bond order as the structure distorts. The fact that net reduction in bond lengths is not observed in either hole- or electron-doped regimes does pose the question why this is.  It could be that the Fe-$X$ disorder somehow enables this enhanced bonding in the Fe-Fe layer. Recall that the structural transition only makes $a_{\rm{tet}}$ non-orthogonal with $b_{\rm{tet}}$, but the magnitudes are still equivalent. Physically this distance is across the diagonal of the Fe square net, that is the next nearest neighbor (nnn) distance.  Moreover, the $X$ site is directly above the midpoint of the nnn interaction, meaning that the nnn distance is directly tied to the Fe-$X$-Fe bond angle. The fact that phosphorus substitution appears to uniquely allow for this separation to decrease through the phase transition suggests that the connection between the phase transition (splitting of nn interactions into long and short interactions) and the nnn contraction is tied to disorder on the pnictogen site. Presumably the local positions of the P sites differ enough from the ideal As site to make these types of distortions energetically favorable on average.  This implies that the local bonding environment of the P atom is different enough from the As atom to warrant future study to elucidate how this substitution affects the underlying interactions and degrees of freedom relating to the structural transition in the BaFe$_2$(As$_{1-x}$P$_x$)$_2$ family.  

\section{Conclusion\label{conclusion}}
Within the sensitivity of these measurements the orthorhombic and magnetic transition in BaFe$_2$(As$_{1-x}$P$_x$)$_2$ occur concurrently as a first-order phase transition. The small amounts of residual peak broadening above the transition are attributed to magnetoelastic fluctuations that are typical of AFM materials above $T_{\rm{N}}$. We observe that the shape of the superconducting dome in BaFe$_2$(As$_{1-x}$P$_x$)$_2$ has a very sharp onset region and is markedly more asymmetric than that observed in the aliovalently-substituted materials.  The suppression of the SDW phase also drops precipitously in this region, disappearing between $x=$ 0.25 and 0.29.  The compositional dependence of the orthorhombic order parameter power-law prefactor follows a linear trend, and extrapolation to zero gives an estimated disappearance of orthorhombicity around $x = 0.279$, in good agreement with the phase line estimation.  This gives a fairly confident estimation of complete suppression of structural order well below the previously postulated QCP around $x$ = 0.33, and is below the maximal $T_{\rm{c}}$ observed, which are usually considered correlated.

It was also shown that there is a lattice anomaly present in the $c/a$ ratio in this isovalently-substituted family that does not appear in either the hole- or electron-doped systems, which may indicate the role of anion site disorder in these matierals. Future studies need to more directly measure the Fe-As and Fe-P separations individually, which is important to more precisely determine the local structure of the Fe$_2$(As,P)$_2$ layer and through that the underlying physics of the system.

\begin{acknowledgments}
The work at the Materials Science Division at Argonne National Laboratory was supported by the U.S. Department of Energy, Office of Science, Materials Sciences and Engineering Division. The part of the research that was conducted at ORNL's High Flux Isotope Reactor and Spallation Neutron Source was sponsored by the Scientific User Facilities Division, Office of Basic Energy Sciences, US Department of Energy. The authors thank A. Huq and P. Whitfield for providing help during experimental collection and analysis.
\end{acknowledgments}

\bibliography{sublibrary}

\end{document}